\documentclass[fleqn,12pt,twoside]{article}
\usepackage{espcrc1}
\usepackage{bm}

 \newcommand{\AmS}{{\protect\the\textfont2
  A\kern-.1667em\lower.5ex\hbox{M}\kern-.125emS}}

\def\He#1{{}^{#1}{\mbox{He}}}
\def\H#1{{}^{#1}{\mbox{H}}}
\def\be{\begin{eqnarray}}
\def\ee{\end{eqnarray}}
\def\nlo#1{\mbox{N$^{#1}$LO}}
\def\vr{{\bm r}}
\def\vA{{\bm A}}
\def\vs{{\bm \sigma}}
\def\dR{{\hat d}^R} \def\calM{{\cal M}}

\title{EFT* for electroweak processes of light nuclei}

\author{Tae-Sun Park\address[MCSD]{School of Physics, Korea Institute for Advanced Study,
Seoul, 130-722 Korea}
\thanks{Invited talk at the 17th International IUPAP Conference on Few-Body
Problems in Physics (FB17), Durham, North Carolina, USA, June 5-10, 2003.}}

\begin{document}

\maketitle

\begin{abstract}
Recently we succeeded to make a reliable EFT prediction in a
totally parameter-free manner for the $S$ factors for the solar
$pp$ and $hep$ processes, $p+p\rightarrow d + e^+ +\nu_e$ and
$\He3+p \rightarrow \He4 + e^+ + \nu_e$\cite{coll}. The strategy
used in there is to embed a highly sophisticated standard nuclear
physics approach (SNPA) exploiting realistic
potentials\cite{carlson,MSVKRB} into an EFT framework, that we
refer to as EFT$*$. Up to next-next-to-leading order in chiral
expansion, it turnes out that there is effectively only one
counter-term relevant to this process, the coefficient of which --
($\hat d^R$) -- has been renormalized to reproduce the
experimental value of the tritium-beta decay. Our study has also
led to  {\em very} accurate EFT calculations on two-body weak
processes that also receive contributions from the $\hat d^R$
term,
$\mu-d$ capture rate\cite{mud}, and $\nu-d$ scattering cross section\cite{nud}.

\end{abstract}

\section{Introduction}
In making theoretical predictions on electroweak processes of
nuclear systems at low energy, we have basically two distinct
approaches: One is the traditional model-dependent approach based
on accurate nuclear potentials and the other is effective field
theory (EFT) approach. The former, that we refer to as ``standard
nuclear physics approach (SNPA)'', has so far scored an enormous
success in wide areas of nuclear physics, achieving in some cases
an accuracy that defies the existing experimental precision. This
suggests that all of the essential ingredients of QCD -- believed
to be the mother theory of strong interactions -- have been
correctly encoded in SNPA. It however suffers from
model-dependence and lack of systematic expansion scheme. In EFT
approach, processes are described consistently and systematically,
and free from the mentioned problems. Respecting these nice
aspects, intensive works based on EFT are recently being made with
impressive successes. For complicated systems with
$A=3,\,4,\,\cdots$, it is technically very hard to apply EFT
directly and not much progress has been made up to date, although
intensive efforts are being made in this direction. Now our
question is, can we construct an EFT that exploits
phenomenological but accurate SNPA wave functions, taking
advantage of the merits of these two processes ? Our answer on
this question is yes, and we have recently developed such a
formalism which we call ``more effective effective field theory
(MEEFT)'' or simply EFT*\cite{npp,BR2001,PKMR,kubodera}. In EFT*,
it is possible to have the consistency and systematic aspect of
EFT with the wave functions obtained accurately in SNPA. In this
talk, I wish to report our developments on EFT*.

To be concrete, we shall discuss the following two solar nuclear
fusion processes: \be pp:&&\ \ \ p+p\rightarrow d + e^+ +\nu_e\,,
\label{pp}
\\
hep:&&\ \ \ p+\He3 \rightarrow \He4 + e^+ + \nu_e\,. \label{hep}
\ee Both figure importantly in the solar neutrino problems. Since
the thermal energy of the interior of the Sun is of the order of
keV, and since no experimental data is available for such
low-energy regimes, one must rely on theory for determining the
astrophysical $S$-factors of the solar nuclear processes. Here we
concentrate on the threshold $S$-factor, $S(0)$, for the reactions
(\ref{pp}) and (\ref{hep}). The necessity of a very accurate
estimate of the threshold $S$-factor for the $pp$ process,
$S_{pp}(0)$, comes from the fact that $pp$ fusion essentially
governs the solar burning rate and the vast majority of the solar
neutrinos come from this reaction. Meanwhile, the $hep$ process is
important in a different context. The $hep$ reaction can produce
the highest-energy solar neutrinos with their spectrum extending
beyond the maximum energy of the ${}^8\mbox{B}$ neutrinos.
Therefore, even though the flux of the $hep$ neutrinos is small,
there can be, at some level, a significant distortion of the
higher end of the ${}^8\mbox{B}$ neutrino spectrum due to the
$hep$ neutrinos. This change can influence the interpretation of
the results of a recent Super-Kamiokande experiment that have
generated many controversies related to neutrino
oscillations~\cite{controversy,monderen}. To address these issues
quantitatively, a reliable estimate of $S_{hep}(0)$ is
indispensable.

Before going further, I would like to explain the difficulty in
making a reliable estimation of $S_{hep}(0)$. First of all, the
leading one-body contribution for the $hep$ process is strongly
suppressed due to the pseudo-orthogonality between initial and
final wave functions. Secondly the main two-body (2B) corrections
to the ``leading" 1B term tend to come with the opposite sign
causing a large cancellation. The 2B terms therefore need to be
calculated with great precision, which is a highly non-trivial
task. Indeed, an accurate evaluation of the $hep$ rate has been a
long-standing challenge in nuclear physics~\cite{controversy}. The
degree of this difficulty may be appreciated by noting that
theoretical estimates of the $hep$ $S$-factor have varied by
orders of magnitude in the literature.

\section{Theory}
The primary amplitudes for both the $pp$ and $hep$ processes are
of the Gamow-Teller (GT) type. Since the single-particle GT
operator is well known at low energy, a major theoretical task is
the accurate estimation of the meson-exchange current (MEC)
contributions. In getting the current operators, we rely on the
heavy-baryon chiral perturbation theory (HB$\chi$PT) with the
Weinberg's power counting rule, which is a well-studied EFT that
has been proven to be quite powerful and successful in describing
low-energy nuclear systems. In HB$\chi$PT, we have pions and
nucleons as pertinent degrees of freedom, with all other massive
degrees of freedom integrated out. The expansion parameter in
HB$\chi$PT is $Q/\Lambda_\chi$, where $Q$ stands for the pion mass
and/or the typical momentum scale of the system, and
$\Lambda_\chi\simeq 1\ \mbox{GeV}$. In our studies, we have
calculated the MEC of the GT operator (space part of the
axial-vector currents) up to \nlo3, where $\nlo{\nu}$ stands for
the order of $(Q/\Lambda_\chi)^\nu$ compared to the leading order
(LO) one-body operator. Up to this order, MEC consists of
one-pion-ranged and contact (zero-ranged) two-body currents,
$\vA_{{\rm 2B}}= \vA^{1\pi} + \vA^\delta$. Three-body and
many-body operators appear only in higher orders. $\vA^{1\pi}$ can
be determined unambiguously, thanks to available accurate $\pi N$
scattering data. $\vA^\delta$ however contains one unknown
parameter, the overall strength of the contact-term contribution,
which we denote by $\dR$, \be \vA^\delta\propto \dR\, \sum_{i<j}
(\tau_i-\tau_j)^{1-i2} (\vs_j\times\vs_j)
\delta_\Lambda^{(3)}(\vr_{ij}), \ee where $\Lambda$ is the cutoff
(which we introduce in the procedure of Fourier transformation
from momentum space to coordinate space), and
$\delta_\Lambda^{(3)}(\vr)$ is the smeared delta-function with the
radius $\simeq 1/\Lambda$.
$\dR$ represents all the heavy degrees of freedom integrated out,
and chiral symmetry (or any other symmetry) does not tell us the
value of it. We can however fix $\dR$ by studying other processes
which are sensitive on it.  In other words, we adjust the value of
$\dR$ to reproduce the experimental data of those other processes,
a procedure which is the {\em renormalization condition} of $\dR$.
This procedure is closely analogous to the EFT approach to
effective nuclear potential backed by renormalization group
equations as explained in ~\cite{doubledecimation}. The power of
the approach is that the constant $\dR$ appears in tritium
$\beta$-decay, $\mu$-capture on a deuteron, and $\nu$--$d$
scattering, as well as in $pp$ and $hep$ processes and hence is
completely fixed. Among them, accurate experimental data is
available for the tritium-beta decay rate, $\Gamma_\beta^t$, which
we use to fix $\dR$.

In EFT*, we evaluate the transition matrix elements by sandwiching
the obtained current operators between phenomenological but
accurate SNPA wave functions. Recalling the fact that, the
short-distance behaviors can be substantially different from model
to model even among modern potentials, one may worry about
model-dependence. To address this question, it is important to
observe that, as was also recently discussed in Refs.
\cite{doubledecimation}, the dependence occurs only at
short-distance, and thus the model-dependence is {\em local}. And
we know that local (short-distance) contribution is well described
(or controlled) in terms of the counter-terms, whose generic form
is $\sum_\nu c_\nu \nabla^\nu \delta_\Lambda^{(3)}(\vr)$. The
$\dR$-term is the leading counter-term.
It means that, for a fixed $\Lambda$, the matrix element of $\delta^{(3)}_\Lambda$,
$\langle \delta^{(3)}_\Lambda\rangle$, may have substantial
model-dependence (due to the model-dependent short-range behavior),
which is, however, compensated by the model-dependence of the value of $\dR$,
to reproduce $\Gamma_\beta^t$. As a result, we can have the {\em model-independent}
theory prediction for the total net amplitude. In other words, EFT* can be different
from EFT only at short-distance, which is to be renormalized away order by order.

\section{Results}
To determine $\dR$ from $\Gamma_\beta^t$, we calculate
$\Gamma_\beta^t$ from the matrix elements of the current operators
evaluated for accurate $A$=3 nuclear wave functions. We employ
here the wave functions obtained in Refs.~\cite{MSVKRB,roccoetal}
using the correlated-hyperspherical-harmonics (CHH)
method~\cite{VKR95,VRK98}. It is obviously important to maintain
consistency between the treatments of the $A$=2, 3 and 4 systems.
We shall use here the same Argonne $v_{18}$ (AV18)
potential~\cite{av18} for all these nuclei. For the $A\ge 3$
systems we add the Urbana-IX (AV18/UIX) three-nucleon
potential~\cite{uix}. Furthermore, we apply the same
regularization method to all the systems in order to control
short-range physics in a consistent manner.


In Table~\ref{tb:tb1}, we have listed the value of $\dR$ and matrix
elements of MEC, $\calM_{\rm 2B}$. The table indicates that,
although the value of $\dR$ is sensitive to $\Lambda$, $\calM_{\rm 2B}$
is amazingly stable against the variation of $\Lambda$ within
the stated range. The $\Lambda$-independence of the physical quantity
$\calM$, which is in conformity with the {\it tenet} of EFT, is a
crucial feature of the present EFT* result.
\begin{table}[t]
\caption{\protect The strength $\dR$ of the contact term and
the two-body GT matrix element, $\calM_{\rm 2B}$, for the $pp$
process calculated for representative values of $\Lambda$.}
\begin{tabular}{|c|c|l|} \hline $\Lambda$ (MeV) & $\dR$ & $\calM_{\rm 2B}$ (fm) \\
\hline 500 & $1.00 \pm 0.07$ & $0.076 - 0.035\ \dR \simeq 0.041 \pm 0.002 $\\
\hline 600 & $1.78 \pm 0.08$ & $0.097 - 0.031\ \dR \simeq 0.042 \pm 0.002$ \\
\hline 800 & $3.90 \pm 0.10$ & $0.129 - 0.022\ \dR \simeq 0.042 \pm 0.002$ \\
\hline \end{tabular} \label{tb:tb1}
\end{table}
Since the $pp$ amplitude is dominated by the well-known one-body LO contribution, the
resulting $S$ factor at threshold, $S_{pp}(0)$, can be predicted
with extreme accuracy,
\be S_{pp}(0) &=&
 3.94\times
  (1 \pm 0.0015 \pm 0.0010)
\ 10^{-25}\ \mbox{MeV-b}.
  \label{S-factor}
\ee Here the first error is due to uncertainties in the input
parameters in the one-body part, while the second error represents
the uncertainties in the two-body part.

In Table~\ref{TabL1A}, we have listed results for the matrix
element of the GT operators for the $hep$ processes in arbitrary
unit.
\begin{table}[b] \caption{\label{TabL1A}\protect Values of
the two-body GT amplitude (in arbitrary unit) for the $hep$
process calculated as functions of the cutoff $\Lambda$. The
individual contributions from the one-body (1B) and two-body (2B)
operators are also listed.}
\begin{tabular}{|c|rrr|} \hline
$\Lambda$ (MeV) & 500 & 600 & 800 \\ \hline
1B                    & $-0.81$ & $-0.81$ & $-0.81$ \\ \hline
2B (without $\dR$)    & $0.93$  & $1.22$  & $1.66$  \\
2B ($\propto \dR$)    & $-0.44$ & $-0.70$ & $-1.07$ \\ \hline
2B-total       & $0.49$  & $0.52$  & $0.59$  \\ \hline
\end{tabular}
\end{table}
%
We see from the table that the variation of the two-body GT amplitude (row labelled ``2B-total'') is $\sim$10 \%
for the range of $\Lambda$ under study. It is also noteworthy
that the variation of the 2B amplitude as a function of $\Lambda$
is reduced by a factor of $\sim$7 by introducing the $\dR$-term
contributions; compare the third and fifth rows (labelled
``2B (without $\dR$)'' and ``2B-total'', respectively) in
Table ~\ref{TabL1A}. As discussed, the $\Lambda$-dependence is
amplified in total (1B $+$ 2B) amplitude due to the substantial
cancellation between 1B and 2B contributions. The resulting $S$-factor
(adding the contributions from non-GT channels) reads
\be S_{hep}(0)=(8.6 \pm 1.3 ) \times \mbox{$10^{-20}$ keV-b}\,,
\label{prediction} \ee
where the ``error" spans the range of the $\Lambda$-dependence
for $\Lambda$=500--800 MeV. This result should be compared to
that obtained by Marcucci {\it et. al.}~\cite{MSVKRB},
$S_{hep}(0)=9.64 \times \mbox{$10^{-20}$ keV-b}$.

\section{Discussions}
The {\it hen} process, $\He3+n \rightarrow \He4 + \gamma$, shares
many features with the {\it hep} process, including the
suppression of the leading one-body contribution due to the
pseudo-orthogonality of the wave functions. Especially, the {\it
hen} process also contains two counter-terms, one in isoscalar and
the one in isovector channel, which can be renormalized by the
magnetic moments of $\He3$ and $\H3$. We note that the MEC of the
$hen$ process starts from \nlo{} coming from unsuppressed
one-pion-exchange diagrams, while that of $hep$ starts from \nlo3.
We are applying exactly the same strategy used in the {\it hep}
process (work in progress with Y-H. Song). Accurate experimental
data is available for the {\it hen} cross section, but so far no
theoretical calculations have succeeded in explaining the data
quantitatively. Thus applying the same EFT technique will provide
a useful check of the validity of EFT* as a bona-fide EFT of QCD
as discussed in Ref.~\cite{BR2001,doubledecimation}.

This talk is based on the work done in collaboration with L.E.
Marcucci, R. Schiavilla, M. Viviani, A. Kievsky, S. Rosati, K.
Kubodera, D.-P. Min,  M. Rho, S. Ando, F. Myhrer and Y.-H. Song,
to all of whom I wish to express my sincere thanks. This work was
supported by Korea Research Foundation
                Grant(KRF-2001-050-D00007).

\end{document}